\documentstyle[12pt,epsfig,a4wide]{article}

\begin{document}

\title{{\bf Statistical Physics and Practical Training of Soft--Committee Machines}}
\author{{\sc M. Ahr$^{1}$, M. Biehl$^{1}$, and R. Urbanczik$^2$}\\
(1) Institut f\"ur Theoretische Physik \\
    Julius-Maximilians-Universit\"at W\"urzburg\\
    Am Hubland, D-97074 W\"urzburg\\
(2) Neural Computing Research Group\\
    Aston University\\ Aston Triangle, Birmingham B4 7ET, U.K.}
\date{}

\maketitle

\begin{abstract}
Equilibrium states of large layered neural networks with differentiable 
activation function and a single, linear output unit are investigated using the 
replica formalism. The quenched free energy of a student network 
with a very large number of hidden units learning a rule of perfectly matching
complexity  is  calculated analytically.  
The system undergoes a first order phase transition from 
unspecialized to specialized student configurations at a 
critical size of the training set. Computer simulations of 
learning by stochastic gradient descent  from a fixed training set 
demonstrate that the equilibrium results describe quantitatively the 
plateau states which occur in practical training procedures at sufficiently
small but finite learning rates. 
\end{abstract}
\medskip

Methods from statistical physics have been applied with great success 
within the theory of learning in adaptive systems. One prominent 
example is the investigation of
feedforward neural networks which are capable of learning an unknown rule
from example data \cite{hertz,bishop}.
  Frequently, 
the training procedure, i.e. the choice of network parameters (weights),
is based on an energy function which measures 
the agreement of the student network 
with the rule in terms of the given example outputs.  
Statistical mechanics techniques can be applied if 
training is interpreted as a stochastic process which leads
to a properly defined thermal equilibrium 
[3-5].
%\cite{kinzelopper,sst,review}.
A particularly interesting topic is that of phase trasitions in this
context, see \cite{kinzel} for a recent review. 
In multilayered neural networks, for example, underlying symmetries
can cause a discontinuous dependence of the success of learning on the 
size of the training set, see e.g.\ 
%\cite{kang,schwarzehertz,schwarze,opper,schottky,robert}. 
[14-19].

In this paper we present the first treatment of learning in fully connected
soft--committee machines by means of the replica method. 
This type of network consists  
of a layer with $K$ hidden units, all of which are connected with the
entire input, and the total output of the net is proportional to the sum of  
their states. 
Previous studies have addressed large soft committees 
$(K\!\to\!\infty)$ with binary weights within the so--called Annealed Approximation
\cite{kang} or networks with finite $K$ in the limit of 
high training temperature \cite{epl}.  

Here, analytical results for the learning of a perfectly realizable rule
at arbitrary training temperatures are derived (for very large $K$)
within a replica symmetric ansatz.
With an increasing size of the training set, the model exhibits a 
first order transition from  unspecialized student configurations
to specialized states with better performance. This transition is 
due to the invariance of the soft--committee output under permutation
of hidden units. 
The same symmetry is known to result in quasi--stationary
plateaus of the learning dynamics in on--line learning from a sequence of independent
training examples 
%\cite{biehlschwarze,saadsolla, brw,vicentecaticha,saadrattray},
[7-11],
see \cite{cambridge} for a recent overview of this framework.
Here, on the contrary, we will consider off--line learning from a fixed, limited
set of examples.     
Furthermore we demonstrate that the statistical physics results,
if interpreted correctly, describe the behavior of practical learning 
prescriptions. To this end we compare our results
with the outcome of a stochastic variant of the well--known
{\sl backpropagation of error\/} algorithm \cite{hertz,bishop,chauvin}. 

We investigate a student--teacher scenario  
where the rule is parametrized as 
\begin{equation} \label{rule} 
\tau(\underline{\xi})  = \,  \frac{1}{\sqrt{K}} \, \sum_{j = 1}^{K} g(y_{j})   
\mbox{~~with~~} y_{j} = \, \frac{1}{\sqrt{N}} 
 \underline{B}_j\cdot \underline{\xi}
\end{equation}
We assume an isotropic teacher with orthonormal
weight vectors: $\underline{B}_j\cdot\underline{B}_k = N  
 \, \delta_{jk}$ for all $j,k$.
The training of a perfectly matching student
with outputs $\sigma(\underline{\xi}) = \sum_{j=1}^K g(x_j) / \sqrt{K}$ 
is considered, 
where the arguments 
$x_j = \underline{J}_j \cdot \underline{\xi}  /\sqrt{N} $ are defined 
through adaptive weights $\underline{J}_j$ 
with $\underline{J}_j^2=N$.
The particular choice of the hidden unit activation function, $g(x) = 
\mbox{erf} (x/\sqrt{2})$,
simplifies the mathematical treatment to a large extent \cite{biehlschwarze,saadsolla,epl}. We 
expect, however, that our results apply qualitatively to a large class of sigmoidal 
functions including the very similar and frequently used hyperbolic tangent.  

Learning is guided by the minimization of the training error 
\begin{equation}  
\label{energy} \mbox{~~~~}
\epsilon_{t}  \, = \, \frac{1}{P} \, 
 H\left(\left\{\underline{J}_j\right\}\right) =  \,
\frac{1}{P} \, \sum_{\mu=1}^{P} \epsilon \left(\left\{\underline{J}_i
 \right\}, \underline{\xi}^{\mu} \right) = \, 
\frac{1}{P}
\sum_{\mu = 1}^{P} \frac{1}{2} 
\left( \sigma(\underline{\xi}^\mu)  - \tau(\underline{\xi}^{\mu})\right)^{2} 
\end{equation}
where $P$ is the number of training examples, which we assume 
to scale like $P = \alpha N K$  with $\alpha = {\cal O}(1)$. 
The extensive quantity $H =P\epsilon_t$ plays the role of an 
energy. The replica formalism for the calculation of the corresponding 
quenched  free energy exploits the identity
$ \left\langle\ln Z\right\rangle = 
\left. \left. \partial\left\langle Z^n\right\rangle \right/ \partial n \right|_{n=0}
$
where $\left\langle\ldots\right\rangle$ denotes an average over the set of random
training examples. 
$Z^{n} $ is equivalent to the partition function of $n$ non-interacting 
copies (labled $a=1,2,\ldots,n$) of the investigated system  and reads:
\begin{equation}
Z^{n} =  \int d\mu (\{ \underline{J}_{i}^{a} \}) \prod_{\mu = 1}^{P} \exp 
\left[{
- \frac{\beta}{2} \sum_{a = 1}^{n} \left({ 
\sigma^{a} \left({\underline{\xi}^{\mu}}\right) - \tau \left({\underline{\xi}^{\mu}}\right)}\right)^{2}}\right].
\end{equation}
Here, the measure $d\mu$ is meant to incorporate the normalization 
$\underline{J}_j^{a \, 2}= N $ of the student vectors.      
We perform the quenched average over all possible sets of independent training
inputs $\underline{\xi}^\mu$, 
the components of which are assumed to be  i.i.d.\  Gaussian random numbers
with mean zero and unit variance. One obtains the following form:
\begin{equation}  
\left\langle{Z^{n}}\right\rangle = \, 
\int d\mu(\{\underline{J}_{i}^{a}\}) \,  e^{- P G_{r}}  \mbox{~~~where~~~}
\label{zethochn} 
G_{r} = \,  - \ln \left\langle  \exp \left[- \frac{\beta}{2} \sum_{a = 1}^{n} 
\left({ \sigma^{a} (\underline{\xi}\,) - 
\tau(\underline{\xi}\,)}\right)^{2}\right] \right\rangle_\xi. \label{gr}
\end{equation} 
Here and in the following $\left\langle \ldots \right\rangle_\xi$ denotes an
average over the randomness contained in a single input vector. 
As the examples are independent, the 
quenched average over the training set factorizes. 

The sample average $G_r$ will only depend on the order parameters
$R_{ij}^a = \underline{J}_i^a\cdot\underline{B}_j / N $~ and 
$Q_{ij}^{ab} = \underline{J}_i^a\cdot\underline{J}_j^b / N$~. 
Similarly the generalization 
error  $ \epsilon_g = \frac{1}{2} \langle (\sigma  - \tau )^2 \rangle_\xi $,
 which measures the success of learning by averaging over arbitrary
inputs is given by \cite{saadsolla}
\begin{equation}  
\label{generror}
\epsilon_g = \, \frac{1}{6} + \frac{1}{K\pi} \sum_{i,j=1}^K \left[
  \mbox{arcsin} \left(\frac{Q^{aa}_{ij}}{2}\right) - 
2 \mbox{arcsin} \left(\frac{R^a_{ij}}{2}\right)
\right].
\end{equation}

In this paper we restrict ourselves to networks with a very large number $K$
of hidden units. Non-trivial results can be obtained in the limit $K\to\infty$
but with $K <\!\!< N$ by assuming that the relevant student configurations
will be site symmetric:

\begin{equation}   
\label{sitesymmetric}
R_{ij}^{a} = \left\{
\begin{array}{cc}
R^{a} & \mbox{if~} i=j \\
S^{a} & \mbox{if~} i\neq j
\end{array},\right. \mbox{~~}
Q_{ij}^{aa} \left\{
\begin{array}{cc}
1 & \mbox{if~} i=j \\
C^{a} & \mbox{if~}  i\neq j
\end{array}, \right.  
 \mbox{and~}
Q_{ij}^{ab} = \left\{
\begin{array}{cc}
q^{ab} & \mbox{if~}  i=j \\
p^{ab} & \mbox{if~}  i\neq j  
\end{array} \right. \mbox{~for~} a\neq b.  \label{site-sym}
\end{equation} 
Here and elsewhere in the paper superscripts $a,b$ label replicas, 
whereas $i$ and $j$ are hidden unit indices.  
The restriction (\ref{site-sym}) allows the system to assume unspecialized
$(R^a = S^a)$ or specialized states $(R^a > S^a)$.
Note that the output of a student will be ${\cal O}(\sqrt{K})$
and thus on a different scale than the output of the teacher
if $C^a$ is on the order of 1. So that the magnitudes of the outputs
match, we assume that the  hidden unit cross overlaps ($C^{a},p^{ab}$ and $S^{a}$)
are on the order of $1/K$. As a consequence of this scaling one may show
\cite{robert} that the joint distribution of $\tau$ and the $\sigma^a$
becomes Gaussian in the large $K$ limit.

In the following we use the notation  $\underline{\sigma} = 
\left({\sigma^{1}, \sigma^{2}, \ldots, \sigma^{n}, \tau}\right)^{\top}$, 
and define
a matrix $\mathbf{\mathsf{B}}$ such that $
\underline{\sigma}^{\top} {\mathbf{\mathsf{B}}} \underline{\sigma} = 
\sum_{a = 1}^{n} \left({\sigma^{a} - \tau}\right)^{2}$.
For large $K$ the Gaussian joint distribution of 
$\underline{\sigma}$ is completely specified through the covariance matrix 
$\mathbf{\mathsf{M}} \, = \left\langle \underline{\sigma} \, \underline{\sigma}^\top \right\rangle$,
the elements of which can be expressed in terms of order parameters. 
Hence one obtains the effective Hamiltonian $G_{r}$, equation (\ref{gr}),
\begin{equation}  
G_r = - \ln 
\left\{\frac{(2\pi)^{\frac{-n-1}{2}}}{\sqrt{\det\mathbf{\mathsf{M}}}}
 \int d^{n\!+\!1}\sigma \,\, \exp \left[{- \frac{1}{2} \underline{\sigma}^{\top} 
\left({\beta \mathbf{\mathsf{B}}  + 
\mathbf{\mathsf{M}}^{-1}}\right) \underline{\sigma}}\right]  \right\}
= \frac{1}{2} \ln \left\{\det \left[{\beta \mathbf{\mathsf{M}} 
 \mathbf{\mathsf{B}}
+ \mathbf{\mathsf{1}}}\right]\right\} \label{gergebnis}
\end{equation}
where the r.h.s. is a function of the  site symmetric
order parameters (\ref{sitesymmetric}).
A saddle point integration gives $1/N \, \ln \left\langle Z^n \right\rangle$ 
as the extremum (w.r.t. $\left\{R_{kl}^a,Q_{kl}^{ab}\right\}$) ~ of ~  
$ \exp[- P \, G_r \, + N \, s] $ \mbox{~~where~~} 
\begin{equation}  \label{entropy} 
\mbox{~~~~~~~} 
 s = \frac{1}{N} \ln \int d\mu(\{\underline{J}_{i}^{a}\}) \prod_{k, l = 1}^{K} 
\prod_{a, b = 1}^{n}
\delta (Q_{kl}^{ab} -  N \underline{J}_{k}^{a}\cdot\underline{J}_{l}^{b}) 
\delta (R_{kl}^{a} -   N\underline{J}_{k}^{a}\cdot\underline{B}_{l}).
\end{equation} 
The entropy term can be calculated  by means of 
a saddle point integration itself after writing the $\delta$-functions in their 
integral representation.  One obtains 
\begin{equation} \label{simpleentropy}
 s = 1/2 \ln ( \det 
\mathbf{\mathsf{C}} ) + \mbox{const.},
\end{equation} 
where $\mathbf{\mathsf{C}}$ is the $[(n + 1)K]$-dimensional 
square matrix  of all
cross- and self-overlaps of (replicated) student and teacher vectors \cite{schwarze,robert}.
In the Appendix we sketch a simpler derivation of this result which 
avoids the saddle point method. 

In order to proceed with the analysis, we make a replica symmetric
ansatz, in addition to site symmetry (\ref{sitesymmetric}):
$R^a\!=\!R,S^a\!=\!S,C^a\!=\!C$ and $q^{ab}\!=\!q,p^{ab}\!=\!p$ for $a\neq b$.
This assumption simplifies the evaluation of the determinants
and allows for a straightforward treatment of the limit $n \to 0$.  
In agreement with the scaling of the hidden unit cross overlaps, we reparametrize:
$S = \widehat{S}/K$, $C = \widehat{C}/(K - 1)$, and $p = \widehat{p}/K$. The
parameters $\Delta = R - S$ and $\delta = q - p$ now measure the degree of 
specialization in the network.
Inserting these in the saddle point equations we find
that the condition $\partial f/\partial \widehat{S} = 0$ can only be satisfied,
if  $\widetilde{C} = K (1 + \widehat{C} - \delta - \widehat{p}) = {\cal O} (1)$. After
eliminating $\widehat{C}$ accordingly, we obtain the free energy as a 
function of variables of order one:
\begin{equation}  
\label{free}
\frac{2\beta F}{N K}  = \alpha \left[{ \frac{\beta (v\!-2w\!+\!1/3)}{1\!+\! 
\beta(u\!-\!v)} + \ln \left[{1\!+\!\beta(u\!-\!v)}\right]}\right] 
 +  \frac{\delta\!-\!\Delta^{2}}{\delta\!-\!1} - \ln (1\!-\!\delta) -
\frac{\delta\!+\!\widehat{p}\!-\!(\Delta\!+\!\widehat{S})^{2}}{\widetilde{C}},
\end{equation}
with $u = 1/3 + \widehat{C}/\pi$,~ $v = [2 \mbox{arcsin}(\delta/2) +
\widehat{p}] / \pi$, and $w=[2 \mbox{arcsin}(\Delta/2) + \widehat{S}] / \pi$.  
Terms of order ${\cal}(1/K)$ have been neglected on the r.h.s. of equation (\ref{free}).

For $\alpha = {\cal O} (1)$, the saddle point equations yield 
two different types of solution: 
an unspecialized, committee symmetric branch with
 $\Delta = \delta = 0$ and specialized solutions with $\Delta,\delta > 0$. 
In the first case we find $\widehat{p} = \widehat{S} = 1$
and $\widehat{C} = 0$, with the generalization error $\epsilon_{g} = 1/3 - 1/\pi$
independent of both $\alpha$ and $\beta$. 
In the specialized case we get $\widehat{C} = 0$, $\widehat{p} = 1 - \delta$ and 
$\widehat{S} = 1 - \Delta$, while $\delta$ and $\Delta$ as functions of
$\alpha$ and $\beta$ can be determined only numerically.

Figure 1 (left) shows the generalization error as a function of 
$\alpha$ for three different values of $\beta$.
The system undergoes a first order phase transition from 
a committee symmetric state ($R=S$) to a specialized solution with $R>S$.
At constant training temperature, a locally stable, specialized
configuration appears at a ($\beta$--dependent) 
value $\alpha_{min}$. 
For $\alpha > \alpha_{glob}(\beta)$,  
the specialized solution becomes globally stable. 
Asymptotically, the corresponding generalization error 
$\epsilon_{g}$ and the training error $\epsilon_{t}$ decay like $1/(\alpha \beta)$
for large $\alpha$.
In contrast to the unspecialized phase, at a given $\alpha$ the 
generalization error always decreases with increasing $\beta$ in the
specialized phase.

It is important to note that an unspecialized configuration with
constant $\epsilon_g$ remains locally stable for all $\alpha$. For a given 
$\beta$ the corresponding training error 
is constant with respect to the size of the training set, initially.
At an additional critical value of $\alpha$, the order parameter
$\delta =q-p$ which measure correlations between students in different 
replicas assumes a non-zero  value, whereas in this phase  $\Delta=R-S$ 
remains zero for all $\alpha$. This transition does not affect the
generalization error but it does cause a first order transition
to a slightly higher value of the training error $\epsilon_t$.
The training error continues to increase and approaches 
its asymptotic value $1/3 - 1/\pi$ while $\delta\to 1$ for $\alpha\to\infty$.
The latter indicates that, asymptotically, a unique set of unspecialized
student weights is chosen in all replicas.
Due to the transition , the  training and generalization error
of the unspecialized configuration 
coincide in the limit $\alpha\to\infty$.

Figure 1 (right) displays $\epsilon_t (\alpha)$ for $\beta =100$, where
the above mentioned phase transition is located at $\alpha \approx 6.18$ where
the training error jumps to a slightly larger value.
The transition within the unspecialized phase occurs at values of $\alpha$
which increase rapidly with the training temperature,  for instance at
$\alpha \approx 139 $ for $\beta=10$ and $\alpha \approx 489$ for $\beta=5$.

Our results parallel the findings of  
\cite{schwarzehertz,schwarze} and  \cite{robert}
for large  multilayer networks with threshold activation functions.
We have found essentially the same qualitative
behavior in the limit of infinite training temperature \cite{epl} and 
by applying the Annealed Approximation. However, the transition 
within the unspecialized phase cannot be identified in these simpler frameworks.
It is further quite possible that even the replica symmmetric 
decription of this transition is incomplete.
For threshold activation functions it  was observed in \cite{robert}
that this transition is affected by replica symmetry breaking,  resulting  
in a lower critical value for $\alpha$ than predicted in replica symmetry 
and changing the nature of the transition from first to second order.
A more detailed discussion of this transition for the present case will 
be given elsewhere \cite{unpubl}.

The limit $\beta\!\to\!\infty$ is of particular interest and corresponds
to potentially error free training with $\epsilon_t =0$ for all $\alpha$. 
Within our replica symmetric ansatz we find for $\beta\to\infty$ that the 
system switches from poor to perfect 
generalization ($\epsilon_{g} = 0$) at $\alpha_{min} =\alpha_{glob} = 1$, where
the number of examples coincides with the number of adjustable weights in the 
network. This is a consequence of the smooth, differentiable nature of 
the input--output relation in this type of network. 
Such a transition to $\epsilon_g = 0$ is not observed in networks with 
threshold activation functions and continuous weights. 
The achievement of perfect generalization observed in networks with binary weights 
is due to a completely different mechanism, i.e.\ a freezing transition in the
discrete configuration space, see e.g. \cite{sst,review,schwarzehertz}. 

\begin{figure}[t]
\begin{center}
\setlength{\unitlength}{1cm}
\begin{picture}(14.5,3.0)(0,0)
\put(-0.5,0){\makebox(7,3.0){
\psfig{file=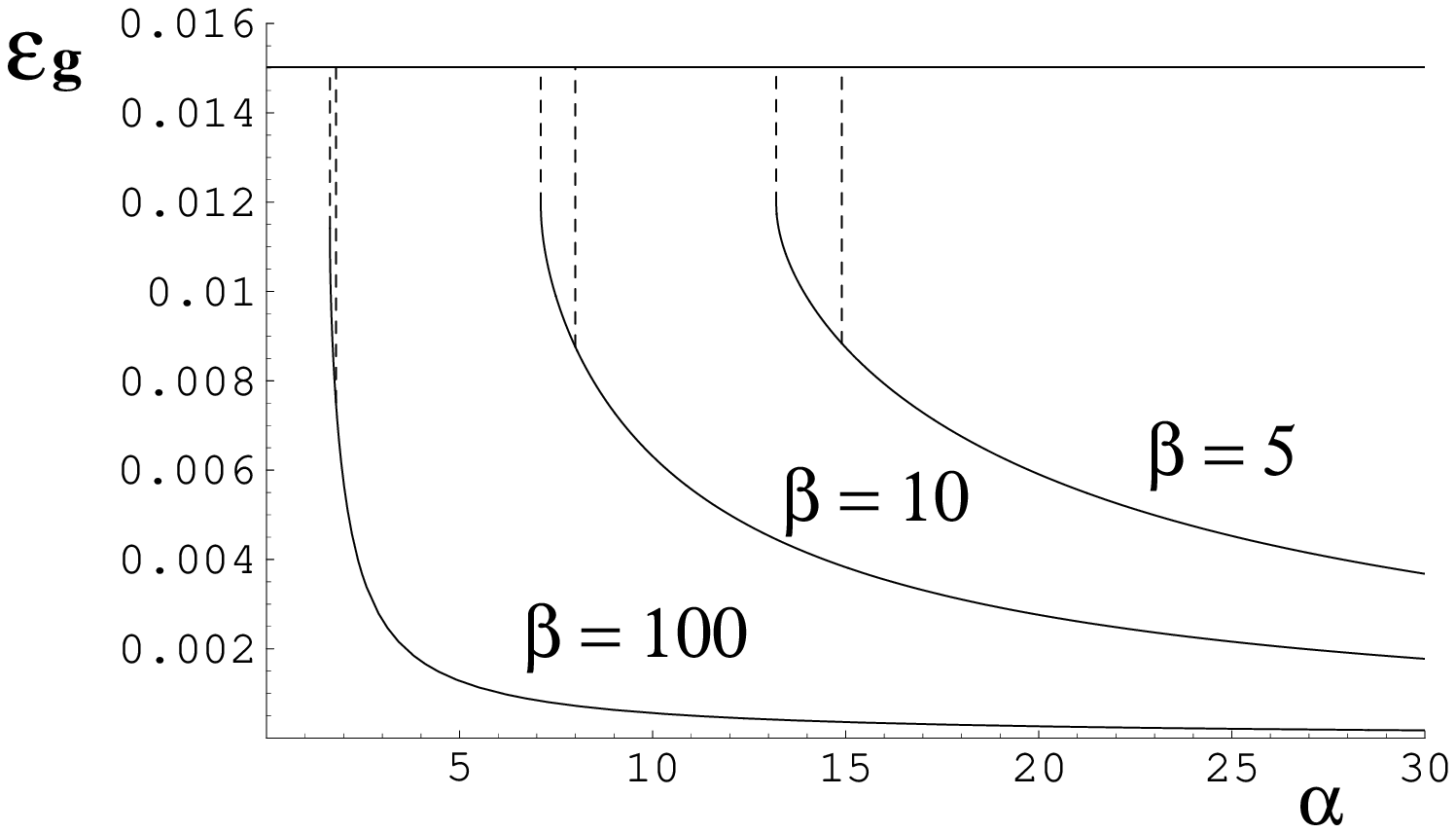,width=6.5cm,angle=270}
}}
\put(7.5,0){\makebox(7,3.0){
\psfig{file=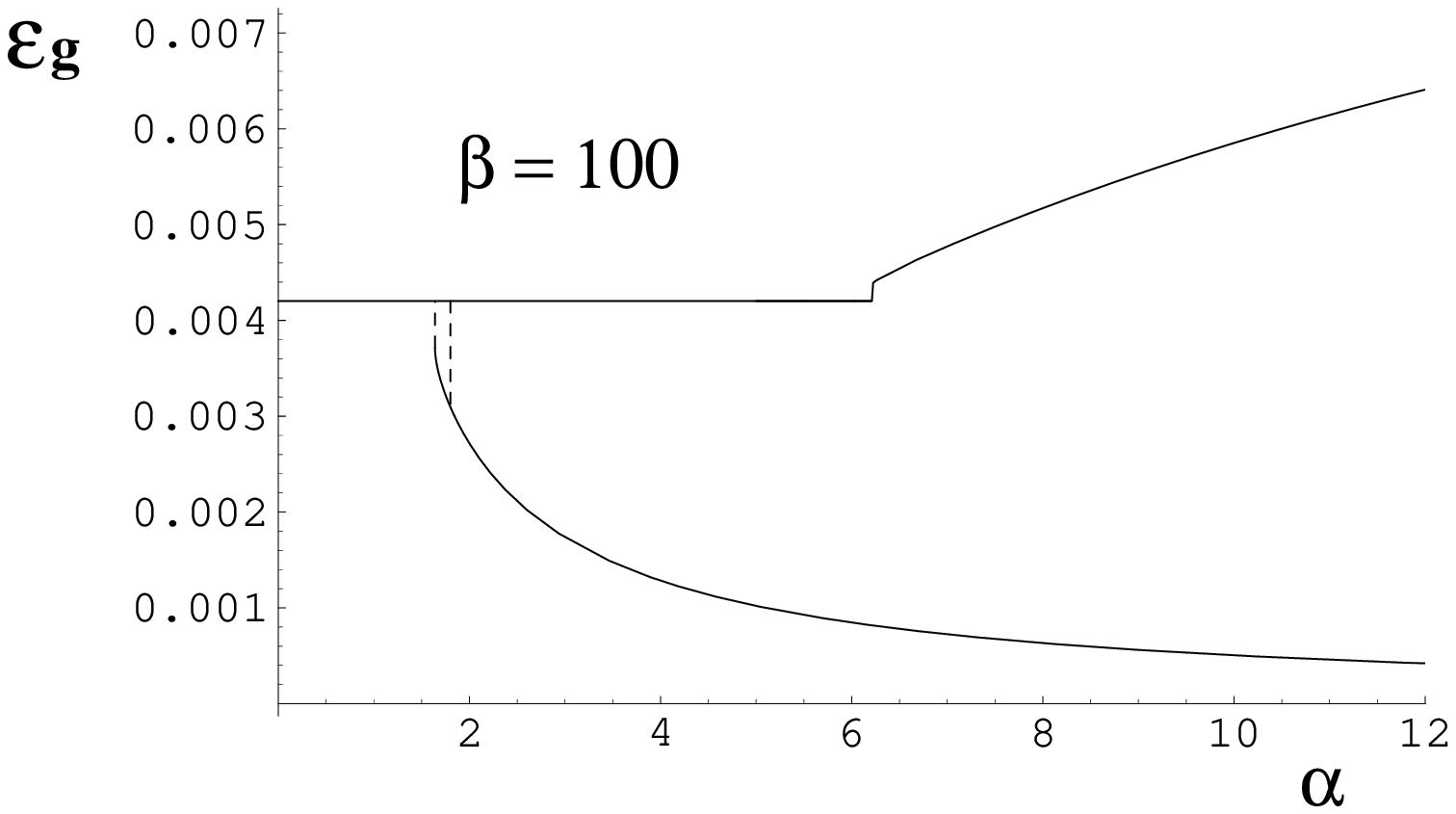,width=6.5cm,angle=270}
}}
\end{picture}
\end{center}
\caption{\label{transition} 
Generalization error and training error as functions of $\alpha = P/(KN)$. \protect\newline
{Left panel:}
$\epsilon_g$ vs. $\alpha$ for three different training temperatures in
the Gibbs ensemble. For each temperature, the leftmost dashed line
indicates the occurance of a locally stable specialized state; the second vertical
 line marks  $\alpha_{glob}$ where it becomes globally stable. \protect\newline
{Right panel:} $\epsilon_t$  vs. $\alpha$ for $\beta=100$. An additional
 (first order) transition occurs at which $\epsilon_t$ begins to increase in the 
 unspecialized solution while $\epsilon_g$ remains constant.
 For $\alpha\to\infty$ the training error 
 approaches the value $\epsilon_t =\epsilon_g = 1/3 - 1/\pi $.  }
\end{figure}
It is of course a crucial question, whether our statistical mechanics 
treatment can give relevant results for practical applications. 
We have followed the standard approach and analysed a heat bath ensemble,
i.e.\ a Gibbs distribution of network configurations. 
One might reproduce the Gibbs density 
in simulations of the learning process by use of
an appropriate Langevin or Monte Carlo dynamics.
However, these prescriptions are out of the question for practical 
applications in the case of continuous weights and differentiable outputs. 
Much faster and more effective methods exist, the most prominent one is certainly 
the so--called {\sl backpropagation of error\/} \cite{hertz,bishop,chauvin}. 

When can we expect the statistical physics results to be relevant for such 
a practical prescription?  Under certain restricting assumptions  one can
show, for instance, that stochastic gradient descent produces a  
stationary distribution which  approximates  a Gibbs density in the limit
of infinitesimally small learning rates. This has been investigated in  detail 
for simple systems in the vicinity of local energy minima 
\cite{hansen,radons,heskes}. 
But heat bath results can be interpreted in a broader context.
Whenever an algorithm yields network configurations with a probability 
which depends exclusively on the training energy, one could in principle analyse 
an appropriate ensemble. All such ensembles, including the heat bath, 
refer to the same microcanonical density. Hence, for fixed energy,
the system chooses among the same set of possible states  with 
equal probablity and the same macroscopic features emerge.    
Stability properties, however, will depend strongly on the considered ensemble
which has to be specified  in order to locate a phase transition, for instance. 

In Figure 2 (left panel) we have plotted  the generalization error {\it vs.\/} 
the corresponding training error at $\alpha = 5$  by eliminating 
$\beta$ in all saddle point solutions (regardless their local or global stability).
Clearly, this dependence could be derived from the microcanonical density as 
well.  According to the above reasoning, the same graph is valid for all
procedures which produce configurations with a purely energy 
dependent probability.

In the following we demonstrate that the backpropagation algorithm 
appears to fulfill this requirement very well for a range of learning rates. 
To this end we have performed simulations of a stochastic version \cite{bishop}
with updates
\begin{equation} \label{learningdyn}
\underline{J}_{i}^{t+1} =  \, \sqrt{N}  \, \, \left. \left( 
\underline{J}_{i}^{t} - \eta \nabla_{\underline{J}_{i}} 
\epsilon (\{\underline{J}_{i}^{t}\}, \underline{\xi}^{\mu(t)}) 
\right)
\,\, \right/ \,\, \left|
\underline{J}_{i}^{t} - \eta \nabla_{\underline{J}_{i}} 
\epsilon (\{\underline{J}_{i}^{t}\}, \underline{\xi}^{\mu(t)}) 
 \right| .
\end{equation}
The current training example $\left\{\underline{\xi}^\mu(t),\tau^\mu\right\}$ 
is drawn randomly 
from the pool of $P=\alpha K N$ independent input-output 
pairs with probability $1/P$ 
at each time step. The learning rate $\eta$ controls the step size of 
this stochastic gradient descent and the weights are normalized 
explicitly. The number of hidden
units was $K=10$ in all simulations shown in Figure \ 2.

In the course of learning  one observes 
quasi--stationary  states in which both $\epsilon_t$ and  $\epsilon_g$
remain almost constant over a large number of updates.  These are reminiscent 
of the {\it plateaus\/} found in on--line training of soft--committees 
[7-9]
%\cite{biehlschwarze,saadsolla,brw}
where each example is presented exactly once. 
We have identified plateaus according to a heuristic criterion in our
simulations and determined the corresponding values of $\epsilon_g$ and 
$\epsilon_t$. Note that several such states can be approached successively while 
learning with a fixed rate $\eta$. 
Details of the simulations will be explained in a forthcoming
publication \cite{unpubl}.

Figure 2 (left) shows the observed pairs of values $(\epsilon_g,\epsilon_t)$ 
for learning rates $0.1 \leq \eta \leq 4.0$.  
Simulation results are in good aggreement 
with the theoretical analysis for a range of finite $\eta$. 
The algorithm favors configurations from 
either one of the two predicted phases, the occurance of states in 
between the specialized and the unspecialized branch  is presumably 
due to the finite size of the system. 
The data with $\epsilon_g$ significantly larger than predicted
correspond  to plateaus found in simulations with relatively large $\eta$.
Figure 2 (right panel) displays the observed values of 
$\epsilon_g$ {\it vs.} $\eta$.  For small enough  learning rates the predicted 
competition of specialized and unspecialized states is confirmed.
For $\eta \widetilde{>} 2$, the value of 
$\epsilon_g$ can deviate significantly from the 
prediction, its  sudden increase at $\eta \approx 5$ is 
reminiscent of the presence of a  critical learning rate in on--line 
learning from a sequence of uncorrelated examples \cite{biehlschwarze,saadsolla}.  

As argued above, the location of a sharp transition from
poor to good generalization cannot be expected to carry over from the 
heat bath to backpropagation results. 
We could not establish a relation between the control parameters $\beta$ and
$\eta$ since the specific density of plateau states as produced by the training 
algorithm is unknown.  Our simulations support, however, the assumption that
it is purely energy dependent for reasonable  $\eta$. The calculation of
student--student overlaps provides further evidence for this hypothesis:
we find the predicted scaling $C \propto 1/K^2 $ for small learning rates, whereas
$ C = {\cal O}(1)$ independent of $K$ for large $\eta$.  Apparently,
stochastic gradient descent with large learning rates prefers, among the states 
of a certain energy, those with highly correlated hidden unit vectors. 

\begin{figure}[t]
\begin{center}
\setlength{\unitlength}{1cm}
\begin{picture}(14.5,3.4)(0,0)
\put(-0.5,0){\makebox(7.0,3.4){
\psfig{file=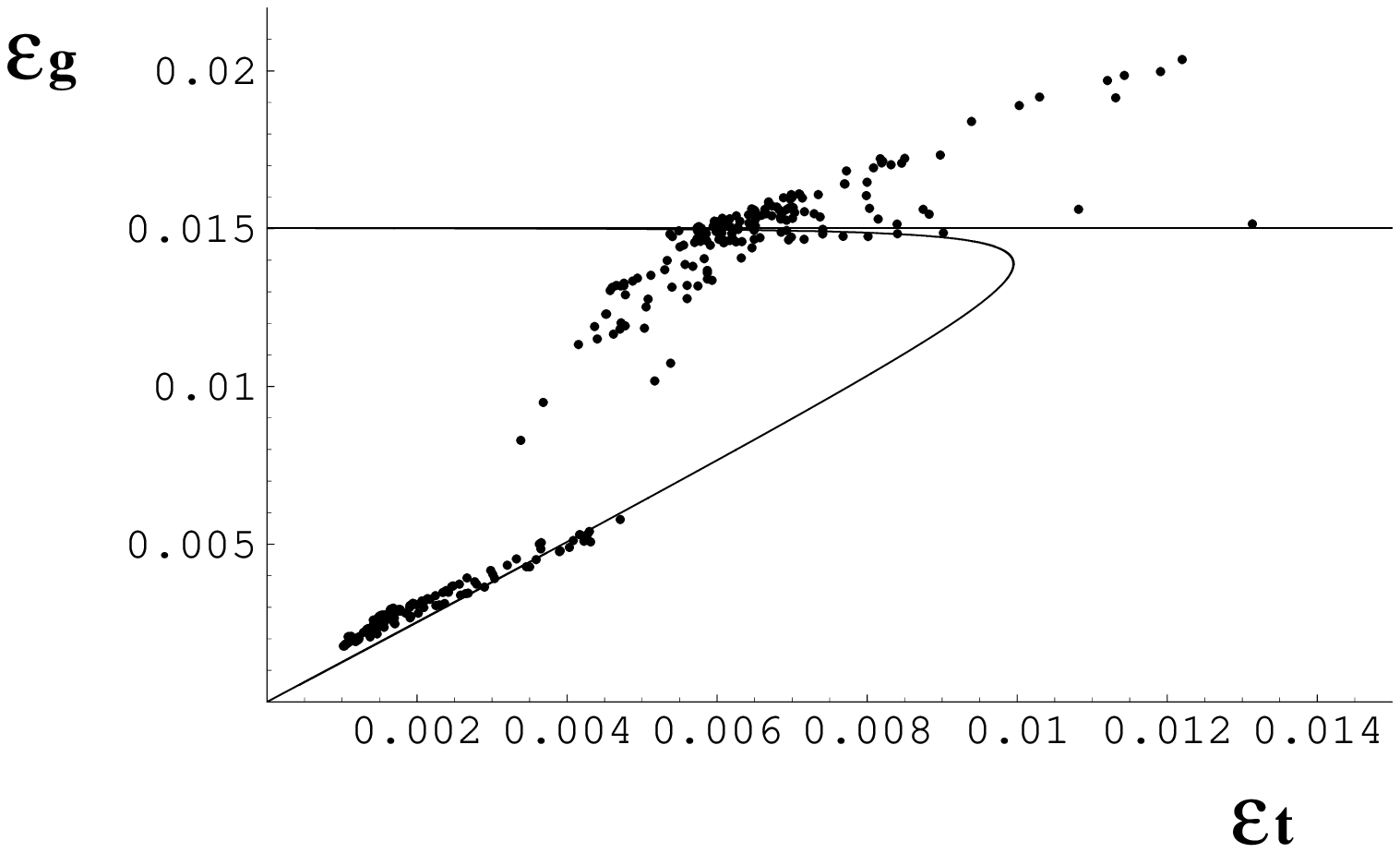,width=6.5cm,angle=270}
}}
\put(7.6,0){\makebox(7.0,3.4){
\psfig{file=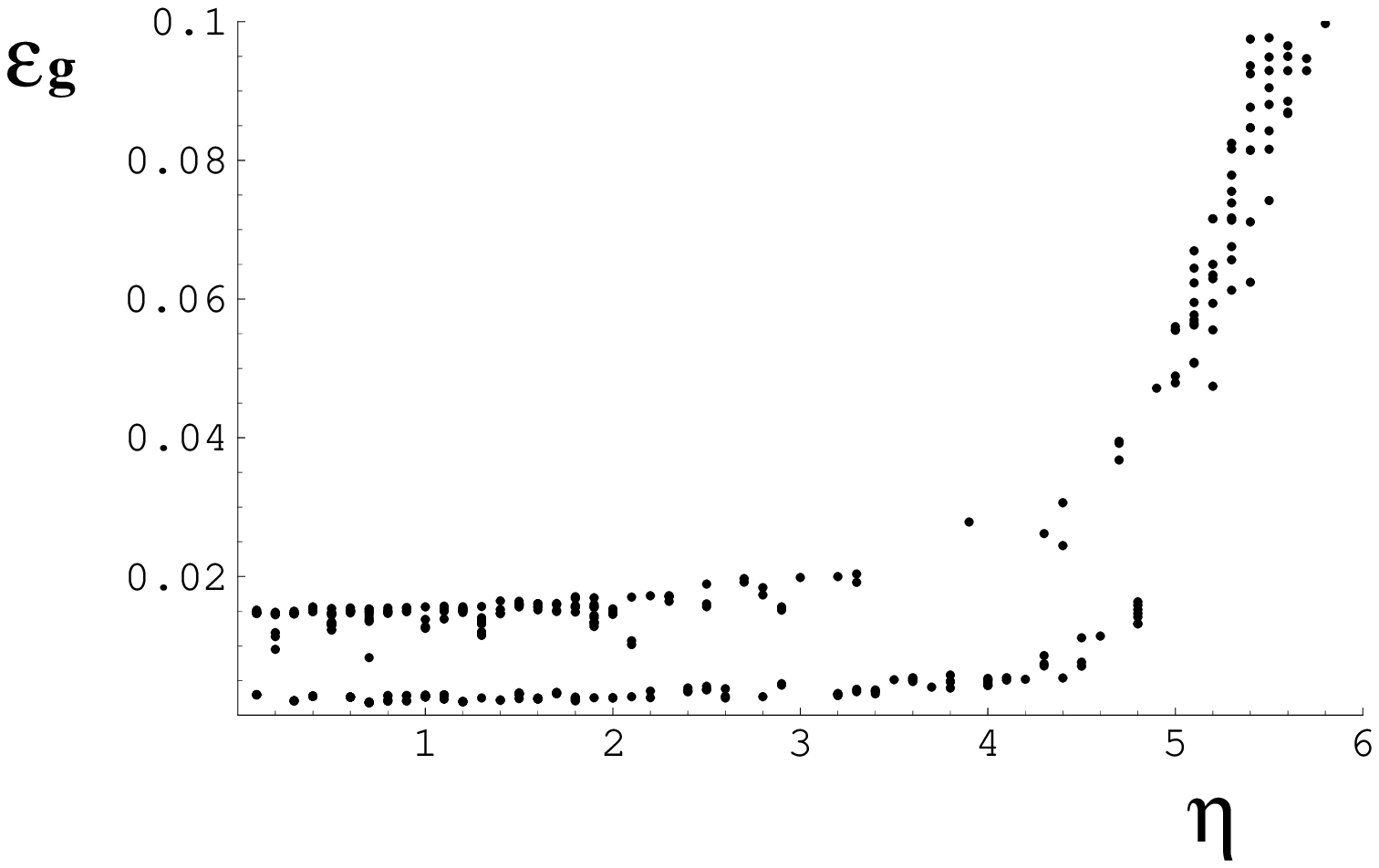,width=6.5cm,angle=270}
}}
\end{picture}
\end{center}
\caption{\label{backprop} Stochastic backpropagation in a system with 
 $N=150$ and $K=10$ at $\alpha = P/(KN)= 5$. 
 Dots represent values found in plateau states of single runs, see
 the description in the text. 
\protect\newline
{Left panel:} Solid lines show $\epsilon_g$ {\it vs.\/} $\epsilon_t$ 
as obtained from the Gibbs ensemble by 
eliminating $\beta$ and disregarding stability
criteria. The dots display the data pairs observed in simulations
with learning rates between $\eta= 0.1$ and $4.0$.\protect\newline
{Right panel:} $\epsilon_g$ as found in  plateau states 
as a function of the learning rate $\eta$. Note that for 
$\eta \widetilde{>} 2$,  $\epsilon_g$ can deviate significantly from the 
prediction. These results contribute to the set of points clearly above 
the horizontal line in the left panel. The generalization error increases 
 drastically for $\eta \widetilde{>} 5$ (not shown in the left panel).
}
\end{figure}

In summary, we have presented an analytic description of  learning
in large soft--com\-mit\-tee machines by means of a replica symmetric 
treatment of the corresponding Gibbs ensemble. 
A characteristic feature of this model is the existence of a first order phase 
transition from  poor to good generalization at a temperature dependent,
critical size of the training set.  In the limit of error free training ($\beta \to
\infty$) the transition is to perfect generalization and occurs at 
$\alpha =1$. 

We expect our results to be relevant for a large class of practical algorithms which
do not favor particular network configurations among those of equal training error.  
Simulations of learning  by stochastic gradient descent with sufficiently 
small but finite learning rates show qualitative and quantitative agreement of plateau
states with the theoretical predictions. This indicates that the considered 
training procedure  provides network configurations with a purely energy 
dependent probability. The latter feature is lost if the learning rate is too large. 

We will provide a more detailed study of stochastic backpropagation in
a forthcoming publication.  Future research will furthermore address learning
from noisy examples, unrealizable rules, and the training of networks with
a finite number of hidden units.  \\[2mm]

\noindent {\bf Acknowledgement:\\} 
We thank G. Reents and E. Schl\"osser for stimulating discussions
and a critical reading of the manuscript.

\ \\

\clearpage
\noindent
{\large \bf Appendix\\}

\noindent
We want to calculate a volume of the form 
\begin{equation}
V({\mathbf{\mathsf{Q}}}) = \int d{\mathbf{\mathsf{J}}} \,\, \mbox{\boldmath$\delta$}
(N{\mathbf{\mathsf{Q}}} -  
{\mathbf{\mathsf{J}}}^\top{\mathbf{\mathsf{J}}}) = 
\int d{\mathbf{\mathsf{J}}} \prod_{a,b=1 (a\leq b)}^n \delta\left(\, NQ_{ab}- \underline{J}^a \cdot \underline{J}^b \,\right)
\end{equation}
 where $\mathbf{\mathsf{Q}}$ is
a symmetric, positive definite $(n,n)$-matrix of overlaps and 
$\mathbf{\mathsf{J}}$ is the $(N,n)$-matrix which
is composed of the $n$ vectors $\underline{J}^a \in \mbox{I}\!\mbox{R}^N$.

For a suitable orthogonal $(n,n)$-matrix $\mathbf{\mathsf{o}}$ and a diagonal $(n,n)$-matrix
$\mathbf{\mathsf{D}}$ one can write $\mathbf{\mathsf{Q}}$ as 
$\mathbf{\mathsf{Q}} = \mathbf{\mathsf{o}}^\top\mathbf{\mathsf{DDo}}$. We now apply the linear
transformation $\mathbf{\mathsf{J}} \rightarrow \mathbf{\mathsf{JDo}}$ 
to the above integral.
Its determinant is $\det \mathbf{\mathsf{D}}^N$ and we obtain 
\begin{eqnarray}
V(\mathbf{\mathsf{Q}})
     &=& \int d{\mathbf{\mathsf{J}}} \,\,\, {\mbox{\boldmath$\delta$}} 
     ({\mathbf{\mathsf{o}}}^\top{\mathbf{\mathsf{D}}}(N{\mathbf{\mathsf{1}}}-
      {\mathbf{\mathsf{J}}}^\top{\mathbf{\mathsf{J}}}){\mathbf{\mathsf{Do}}}) 
     \,\,\, \det {\mathbf{\mathsf{D}}}^N\;.  \label{transJ}
\end{eqnarray}
The Fourier representation of the $\delta$-function yields
\begin{equation}
{\mbox{\boldmath$\delta$}}({\mathbf{\mathsf{o}}}^\top {\mathbf{\mathsf{D}}}
(N{\mathbf{\mathsf{1}}}-{\mathbf{\mathsf{J}}}^\top{\mathbf{\mathsf{J}}})
 {\mathbf{\mathsf{Do}}}) = C_n \int d\hat{{\mathbf{\mathsf{Q}}}} 
 \,\, \exp\left(\, i \, \mbox{\rm Tr} \left[ \hat{{\mathbf{\mathsf{Q}}}}
{\mathbf{\mathsf{o}}}^\top{\mathbf{\mathsf{D}}}(N{\mathbf{\mathsf{1}}}-
{\mathbf{\mathsf{J}}}^\top{\mathbf{\mathsf{J}}}){\mathbf{\mathsf{Do}}}\right] \right)
\,.
\end{equation}
The integration runs over symmetric $(n,n)$-matrices and 
$C_n = (2\pi)^{-n(n+1)/2}2^{n(n-1)/2}$, where the second factor arises
from the fact that the off-diagonal elements are counted twice in the
trace.
Using  
$$ \mbox{Tr}\left[\,\hat{{\mathbf{\mathsf{Q}}}}{\mathbf{\mathsf{o}}}^\top
{\mathbf{\mathsf{D}}}(N{\mathbf{\mathsf{1}}}-{\mathbf{\mathsf{J}}}^\top
{\mathbf{\mathsf{J}}}){\mathbf{\mathsf{Do}}}\, \right] \,\,  = \,\,  \mbox{Tr}\left[\,
{\mathbf{\mathsf{Do}}}\hat{{\mathbf{\mathsf{Q}}}}{\mathbf{\mathsf{o}}}^\top
{\mathbf{\mathsf{D}}}(N{\mathbf{\mathsf{1}}}-
{\mathbf{\mathsf{J}}}^\top {\mathbf{\mathsf{J}}}) \, \right]$$ 
and 
transforming 
$\hat{{\mathbf{\mathsf{Q}}}}$ via 
$\hat{{\mathbf{\mathsf{Q}}}} \to  {\mathbf{\mathsf{o}}}^\top
{\mathbf{\mathsf{D}}}^{-1} \hat{{\mathbf{\mathsf{Q}}}}
{\mathbf{\mathsf{D}}}^{-1} {\mathbf{\mathsf{o}}} $ yields

\begin{eqnarray} {\mbox{\boldmath$\delta$}}
( {\mathbf{\mathsf{o}}}^\top{\mathbf{\mathsf{D}}} (N{\mathbf{\mathsf{1}}}-
{\mathbf{\mathsf{J}}}^\top{\mathbf{\mathsf{J}}}
{\mathbf{\mathsf{Do}}}) & =  & \,\,
C_n \det {\mathbf{\mathsf{D}}}^{-n-1} \int \, d\hat{{\mathbf{\mathsf{Q}}}} 
\,\, \exp\left(\,i\, \mbox{Tr} \left[
\hat{{\mathbf{\mathsf{Q}}}}(
N{\mathbf{\mathsf{1}}}-
{\mathbf{\mathsf{J}}}^\top{\mathbf{\mathsf{J}}})
\right])\right) \nonumber \\
                  & = & C_n \det {\mathbf{\mathsf{D}}}^{-n-1} 
{\mbox{\boldmath$\delta$}}
(N{\mathbf{\mathsf{1}}}-{\mathbf{\mathsf{J}}}^\top{\mathbf{\mathsf{J}}})
\end{eqnarray}
and thus $V({\mathbf{\mathsf{Q}}}) = \det {\mathbf{\mathsf{D}}}^{N-n-1} 
V(N{\mathbf{\mathsf{1}}})$.
Now $V(N{\mathbf{\mathsf{1}}})$ is just a normalization
constant and of course $\det {\mathbf{\mathsf{D}}}^2 = \det {\mathbf{\mathsf{Q}}}$.
Hence, in the limit $N\to\infty$ with $n$ of order one, one obtains
$$
\frac{1}{N} \ln  V({\mathbf{\mathsf{Q}}}) \,\, = \,\,  \frac{1}{2} \, \ln \det   
{\mathbf{\mathsf{Q}}} + {\cal{O}}(1). 
$$

The case where one considers an additional $(N,m)$-Matrix ${\mathbf{\mathsf{B}}}$
of $m$ teacher vectors
and wants to evaluate   $\int \, d{\mathbf{\mathsf{J}}}\,\,
{\mbox{\boldmath$\delta$}} (N 
{\mathbf{\mathsf{Q}}} - 
{\mathbf{\mathsf{J}}}^\top
{\mathbf{\mathsf{J}}}) 
\, {\mbox{\boldmath$\delta$}} (N 
{\mathbf{\mathsf{R}}}-{\mathbf{\mathsf{J}}}^\top 
{\mathbf{\mathsf{B}}}) 
\, $ reduces to the above consideration by noting that the integral will 
not depend on the choice of 
${\mathbf{\mathsf{B}}}$, as long as the matrix of teacher overlaps
${\mathbf{\mathsf{T}}}= {\mathbf{\mathsf{B}}}^\top{\mathbf{\mathsf{B}}}/N $
is held fixed. Thus, one may in addition integrate over 
all ${\mathbf{\mathsf{B}}}$ which have correlation matrix ${\mathbf{\mathsf{T}}}$.

For the system of $K$ teacher vectors and $nK$  replicated
students we define the $(n+1)K$--dimensional square  matrix
of overlaps 
$$
{\mathbf{\mathsf{C}}} = \left( \begin{array}{ll}  
{\mathbf{\mathsf{Q}}} & {\mathbf{\mathsf{R}}} \\ 
{\mathbf{\mathsf{R}}^\top} & {\mathbf{\mathsf{T}}}. \\ 
\end{array} \right)
$$
for which the  above result yields equation (\ref{simpleentropy}).

\end{document}